\documentclass[epj,referee]{svjour}
\usepackage{graphics}
\usepackage{graphics}
\begin{document}
\sloppy
\title
{Lattice bosons in quartic confinement}
\author{R. Ramakumar\inst{1}\thanks{e-mail: rkumar@physics.du.ac.in} 
and A. N. Das\inst{2}}
\institute{Department of Physics and Astrophysics, University of Delhi,
Delhi-110007, India \and Theoretical Condensed Matter Physics Division, 
Saha Institute of Nuclear Physics,
1/AF Bidhannagar, Kolkata-700064, India} 
\date{12 January 2008}
\abstract
{  
  We present a theoretical study of bose condensation
of non-interacting bosons in finite lattices in quartic potentials in
one, two, and three dimensions.  We investigate dimensionality 
effects and quartic potential effects on single boson density of 
energy states, condensation temperature, condensate fraction, 
and specific heat. The results obtained are compared with 
corresponding results for lattice bosons in harmonic traps.
} 
\maketitle
\section{Introduction}
\label{sec1}
Over last few years,
bosons and fermions in optical lattices have emerged as important 
controllable systems for investigations into several properties of quantum 
many-particle systems\cite{bloch}. 
These are clean systems in which experimentalists have achieved great
control over a wide range of particle numbers, particle hopping, 
and strength and sign of inter-particle  interactions.
Experimental groups have conducted extensive studies 
\cite{greiner,stof,schori,paredes,kohl,campbell,folling}
of several properties of many-boson systems in one, two, and 
three dimensional optical lattices. The many-boson system in these 
experiments is under the combined influence of the periodic lattice
potential and an overall confining harmonic potential. Many 
theoretical studies\cite{jaksch,zwer,rey,pupilo1,wessel,pollet,giampaolo,demarco,pupilo2,wild,murg,ramdassil} of such lattice bosons in harmonic confinement 
have appeared in recent years.
Unlike the case of lattice bosons in harmonic confinement,
the effect of {\em anharmonic} potentials on the properties of
lattice bosons have begun to be explored only recently\cite{gygi}.
In that work Gygi and collaborators studied zero temperature
properties of strongly interacting lattice bosons in a quartic trap.
It has been suggested\cite{gygi} that it is experimentally possible to create
an optical lattice in a quartic trap employing a combination of
red-detuned and blue-detuned Gaussian laser beams.
Finite temperature properties of lattice bosons in a quartic
trap is of considerable interest in this context.
\par
In this paper, we present a theoretical study of lattice bosons in
a quartic potential in one, two, and three dimensions (1d, 2d, and 3d). 
We consider non-interacting  bosons in a periodic lattice 
in 1d in a 1d quartic potential, a square lattice in 
a 2d quartic potential,
and a cubic lattice in a 3d quartic potential. 
We study the effects of the  potential on
one-boson density of energy states (DOS) and the temperature
dependence of ground state occupancy and specific-heat.
We compare the results obtained
for lattice bosons in quartic traps to the corresponding results
for lattice bosons in harmonic  traps\cite{ramdassil}. 
This work is presented in the
Sections 2 and 3, and conclusions are given in Section 4.
In the work presented in the following sections, we have not
included the effect of boson-boson interaction ($U$). Our results
would approximately also hold in the weak interaction regime ($U\,<<\,t$,
where $t$ is the boson hopping energy) where the interaction induced
depletion effects are not significant\cite{ramdas}. A weakly 
interacting regime may be achieved by adjusting the lattice potential
depth to a low value as has already been done for lattice bosons
in quadratic traps\cite{greiner,schori}. 
\section{Model and method}
\label{sec2}
In this Section, we give a brief presentation of the  model of the
system and the method followed in the calculations of various
properties presented in later sections.  The Hamiltonian of the many-boson
system we consider is
\begin{equation}
H=-t \sum_{<ij>}\left(c^{\dag}_{i}c_{j}+c^{\dag}_{j}c_{i}\right)
  +\sum_{i}V(i)n_{i}-\mu\sum_{i}n_{i}\,,
\end{equation}
where $t$ is the kinetic energy gain when a boson hop
from site $i$ to its nearest neighbor site $j$ in the optical lattice,
$c^{\dag}_{i}$ is the boson creation operator, $V(i)$
is the quartic potential at site $i$, $n_{i}=c^{\dag}_{i}c_{i}$
the boson number operator, and $\mu$ the chemical potential.
The forms of the quartic potentials used are:
$V(i)$ = $Q\,x_{i}^4$ in 1d,
$V(i)$ = $Q\,(x_{i}^4+y_{i}^4)$ in 2d, and
$V(i)$ = $Q\,(x_{i}^4+y_{i}^4+z_{i}^4)$ in 3d.
We first obtain the matrix representation of the system
Hamiltonian in a site basis.
We numerically diagonalize it to obtain
energy levels of a lattice boson.
We have used open boundary conditions.
We have chosen lattice sizes large enough
so that finite size effects are absent in the results presented in the next
section. The lattice sizes were fixed by finding the  lattice
size beyond which results remain unchanged with further increase
of lattice size. This depends on the magnitude  of the quartic 
potential strength since it decides the spread of the boson
distribution in the lattice for a given value of $t$ and
temperature. The energy levels ($E_i$) obtained for
a boson in  these large lattices are used in calculations of the
DOS, ground state occupancy, and the specific heat. 
The chemical potential and boson populations
in the various energy levels are calculated using
the boson number equation: $N\,=\,\sum_{i=0}^{m}N(E_{i})$,
where $E_0$ and  $E_m$ are the lowest and the highest 
single boson energy levels and 
$N(E_{i})=1/[exp{[\beta\,(E_{i}-\mu)]}-1]$ in which 
$\beta\,=\,1/k_{B}T$ with $k_{B}$ the Boltzmann
constant and $T$ the temperature. 
The specific heat is calculated
from the temperature derivative of 
total energy ($ E_{tot}=\sum_{i=0}^{m}N(E_{i})E_{i}$).
All energies are measured in units of $t$.
\section{Results and Discussion}
\label{sec3}
\subsection{One-boson density of energy states}
In this section, in Figs. 1-4, we present our results on
the one-boson density of energy states (DOS) for a
lattice bosons in a quartic trap and compare with the
corresponding DOS for a lattice boson in
a harmonic trap for which: $V(i)=K(x_{i}^2+y_{i}^2+z_{i}^2)$. 
In Fig. 1, we have exhibited the DOS for
bosons in optical lattices with harmonic and quartic confinement 
potentials for several values of $q\,=\,Qa^4$ and $k\,=\,Ka^2$, where $a$ is
lattice constant.
We find that the confining potential has a significant effect on the
DOS in both cases. 
In the case of quartic confining potential (Fig. 1(a)), the  
divergence in DOS at the band edges is found to be  suppressed and 
eventually destroyed with increasing strength of the potential.
Further, the quartic potential spreads the DOS over a wide
energy scale. It is also to be noted that for small potential
strengths shown in the figure, the low energy part of the DOS
continues to show similarity to the case of the DOS of lattice
bosons without confinement.  In comparison, in the case of harmonic 
potential (Fig. 1(b)), divergence of the DOS at the lower edge 
of the band is destroyed while the
one at higher edge is significantly suppressed even for
a weak harmonic potential, and with increasing strength the DOS
is flattened to a wide energy scale\cite{hooley}. 
In Fig. 2, we have shown our results on the DOS of a boson 
in a 2d square lattice in a 2d quartic potential (Fig. 2(a))
and the corresponding results for the harmonic case(Fig. 2(b)).
In the quartic potential case, we find that the van-Hove
singularity is strongly suppressed by the confinement potential.
In comparison, the Van Hove singularity is destroyed 
by the harmonic potential\cite{hooley}. In both cases the confining potentials
spread the DOS compared to the pure lattice case.  Fig. 3 shows the
DOS of a boson in a cubic lattice in a quartic potential. 
Increasing the quartic potential strength clearly has a strong
effect on the DOS. Similar results are obtained for the harmonic
potential case as well, as shown in Fig. 4. The dotted lines
in Fig. 4 are the single particle DOS for an infinite 
dimensional hypercubic lattice whose DOS is\cite{metzner}: 
$\rho(E) = exp[-E^{2}/(2 t^2)]/\sqrt{2 \pi t^2}$.
On comparing the changes brought about by the harmonic potential in 1d, 2d,
and 3d, we notice an approximate dimensionality crossover
in the DOS for small $k$. 
For a 1d lattice with harmonic potential, the DOS has a finite value
at the lower band edge and a weak singularity well inside the band,
which are characteristics of the DOS of a 2d lattice in the absence
of confining potential. For a 2d lattice with harmonic potential, the DOS
almost vanishes at the lower band edge and has a flat region in the middle
part of the band, which are characteristics of the DOS of a 3d lattice
in the absence of any confining potential.
Finally, the DOS of a 3d lattice with harmonic potential 
is found to be close to that of an infinite dimensional hypercubic lattice. 
Hence in DOS, the dimensionality crossovers seen are:
$1d\,(k\neq 0)\rightarrow 2d\,(k=0)$, $2d\,(k\neq 0)\rightarrow 3d\,(k=0)$,
 and  $3d\,(k\neq 0)\rightarrow \infty d\,(k=0)$. In comparison,
such dimensionality crossovers are not found in the case of
lattice bosons in quartic traps.
\subsection{
Ground state occupancy and specific heat}
In this section we present our results  on condensate fraction,
condensation temperature, and specific heat of lattice bosons in  
quartic potentials.  A comparison of the temperature 
dependence of the fractional ground state occupancy for 1d lattice
bosons in a 1d harmonic trap and in a 1d quadratic trap is
shown in Fig. 5 where we have plotted the variation of $N_0/N$
with $T/T_0$. Here $N_0$ is the boson population in the lowest energy level
and $T_0$ is determined by setting $N_{0}=0$ and $\mu =E_{0}$ in the number
equation (i.e., by solving $N = \sum_{i=1}^{i_{m}}
1/[exp(E_{i}-E_{0})/k_{B}T_{0})-1]$). 
We note that the dependence of fractional ground state occupancy on scaled
temperature ($T/T_0$) is nearly independent of the strengths of the 
potentials when lattices are large enough that finite size
effects are absent. The lattice sizes and the strengths of potentials
we have used satisfy this condition.
The dependence of $T_0$ on the strength of the quartic potential 
presented in Fig. 6
shows a fast increase for small values of $q$ and a monotonic
increase for larger $q$. The boson number dependence of $T_0$ is
found to be linear (not shown) similar to lattice bosons 
in a harmonic potential\cite{ramdassil}. In Fig. 7, we have shown the dependence
of the specific heat on scaled temperature. Increasing the the
strength of quartic potential suppress $C_v$ except in the
low temperature regime. In the low temperature range, the $C_v$
is nearly independent of the potential strength. This can
be qualitatively understood if we regard the DOS in Fig. 1(a).
At low temperatures the bosons are in low energy states.
The DOS plots implies that change in the shape of the
band bottom is not significant for small $q$.
Hence for small $q$, the $C_v$ curves for different $q$
values are very close to each other in the low temperature range.
\par
In Figs. 8-10, we have presented our results for bosons in
2d square lattices in a 2d quartic potential. Compared to the
1d case, the condensate
fraction is found to increase faster with decreasing temperature
in the temperature range below $T_0$. The temperature variation
of $N_{0}/N$ is seen to be close to the case of pure lattice
bosons compared to lattice bosons in a harmonic trap.
This feature is also found in the 3d case discussed later. 
The results of the quartic confinement case is closer to the
pure lattice case since, in the central region of the 
lattice, the quartic  potential is shallower compared to the 
harmonic confining potential. In the low energy range, the DOS of 
a lattice with quartic potential is closer to that of a pure 
lattice compared to the DOS of a lattice with harmonic confining potential.
Increase in dimensionality leads to smaller
number of bosons in the ground state for $T\geq T_0$. Fig. 9 shows 
the dependence of condensation temperature on the strength of
the quartic potential. While the $k$ dependence is similar 
to that found in 1d, the magnitude of $T_0$ is much lower in 2d
compared to that in 1d for the same value of $q$.
Fig. 10 shows the dependence of specific 
heat on temperature for various strengths of the quartic trap
potential. Unlike in 1d, the $C_v$ is found to have a peak
near the condensation temperature. In the intermediate to
high temperature range, the specific heat is found to 
be suppressed with increasing potential strength. 
Low temperature $C_v$ shows a slight enhancement with
increasing $q$. When compared with the low temperature  $C_v$ of lattice
bosons in harmonic traps which has a $(T/T_0)^2$ dependence, 
we find that this low temperature part for the quartic trap has approximately 
a $(T/T_0)^{1.7}$ dependence as shown in Fig. 11. 
This can be understood when we consider
the DOS of a lattice boson in quartic and harmonic traps.
We first recall that the temperature dependence of the specific heat
of bosons goes as $(T/T_0)^\alpha$ for a $E^{(\alpha-1)}$ dependence
of the DOS\cite{pethick}. Now, we find that the low energy part of the DOS
of lattice bosons in a quartic trap can be fitted with a value
of $\alpha \approx 0.8-0.9$,  as a consequence of which the specific heat
exponent is less than 2. For 2d bosons in a 2d harmonic trap,
low energy part of DOS has a linear $E$ dependence leading to
a $(T/T_0)^2$ dependence of $C_v$. 
\par
Now we discuss our results on 3d lattice bosons in a 3d quartic trap.
The temperature dependence of the condensate fraction, shown in
Fig. 12, shows that increased dimensionality makes the condensation
sharper. 
The Bose condensation is favored
by small values of the DOS at the bottom of the energy band.
With increasing dimensionality, the DOS decreases at and near
the bottom of the band and this leads to the sharpness of
the condensation.
Similar to 1d and 2d results, the growth of the 
condensate fraction with temperature is found to lie intermediate 
between the lattice bosons in a harmonic trap and pure lattice bosons.
Clearly, if one is looking for phases of bosons in optical lattices
with minimum intervention from the overall confining potential, 
it would be better to use a quartic confinement rather than a
harmonic one. Fig. 13 contains our results on quartic potential
strength dependence on the condensation temperature. These results
are qualitatively similar to 1d and 2d, and quantitatively similar
to 2d than 1d. The temperature dependence of $C_v$ is shown
in Fig. 14 for various strengths of the quartic potential.
Increased dimensionality makes the peak at the condensation
temperature sharper. Similar to 1d and 2d, increasing the
strength of the potential leads to suppression of $C_v$ 
except in the low temperature region. The low temperature
specific heat has a $(T/T_0)^{2.65}$ dependence as shown in Fig. 11.
For lattice bosons in a harmonic trap, the low temperature
$C_v$ has a $(T/T_0)^{3}$  dependence. This difference originates
from the difference in the DOS for these two cases which, in the
low temperature range, shows a
$E^{(\alpha-1)}$  dependence with $\alpha = 3$ for lattice bosons
in a harmonic trap and $\alpha \approx 2.5-2.6$ for the quartic trap. 
\section{Conclusions}
\label{sec4}
In this paper, we presented results of our calculations of
single boson density of energy states, condensate fraction,
condensation temperature, and specific heat of bosons
in some one, two, and three dimensional periodic lattices in
a quartic potential. Wherever possible, the results obtained
are compared with the corresponding results for lattice bosons
in harmonic traps. In one dimension, we find that the DOS of a lattice boson
in a quartic potential continues to retain its pure lattice
form for small potential strengths unlike the case of harmonic 
trap. In two dimensions, the Van Hove singularity is suppressed but not
eliminated for small quartic potential strengths in contrast to the 
effect of a harmonic potential. In three dimensions, the quartic and
harmonic potential are found to destroy 
flat regions of the DOS of a pure lattice. In addition, for
lattice bosons in a harmonic potential, one finds a dimensionality
crossover in the DOS, which is not found in the case of a quartic potential.
The temperature variation of condensate fraction, condensation temperature,
and specific heat is found to be intermediate between those
of lattice bosons in a harmonic potential and pure lattice bosons.
\section*{Acknowledgments}
RRK thanks Professor Bikash Sinha, Director, SINP and
Professor Bikas Chakrabarti, Head, TCMP Division, SINP
for hospitality at SINP. RRK also thanks
Professor Helmut Katzgraber, ETH, Zurich
for drawing his attention to bosons in quartic traps.
We thank Dr. S. Sil, Visva Bharati, Santiniketan for help in computation.
{}
\newpage
\begin{figure}
\resizebox*{3.1in}{4.5in}{\rotatebox{270}{\includegraphics{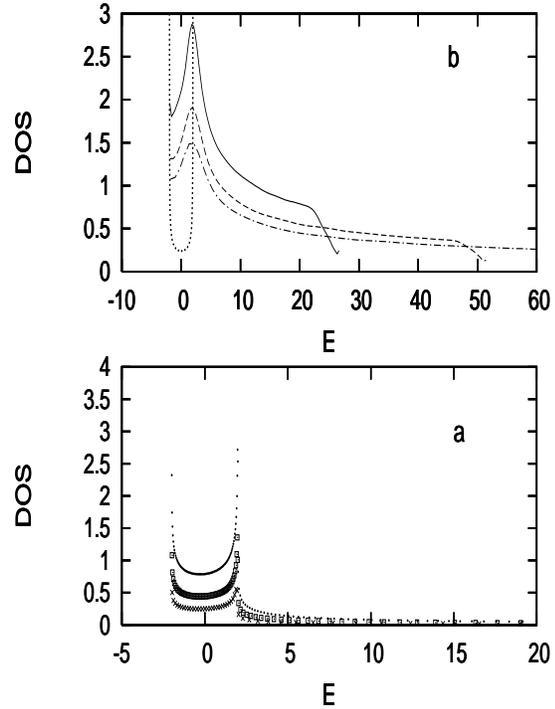}}}
\caption[]{
Density Of States (DOS) of a boson in a one-dimensional
periodic lattice of size 1000 in quartic (a)  and harmonic (b) potentials.
In bottom panel (a): $q\,=\,0.25\times 10^{-6}$
(dots), $0.25\times 10^{-5}$ (squares), and $0.25\times 10^{-4}$ (crosses).
In top panel (b): $k\,=\,0$ (dots), 0.0001 (solid), 0.0002 (dashes),
and 0.0003 (dash-dot). All energies are measured in units of $t$. 
}
\end{figure}
\begin{figure}
\resizebox*{3.1in}{4.5in}{\rotatebox{270}{\includegraphics{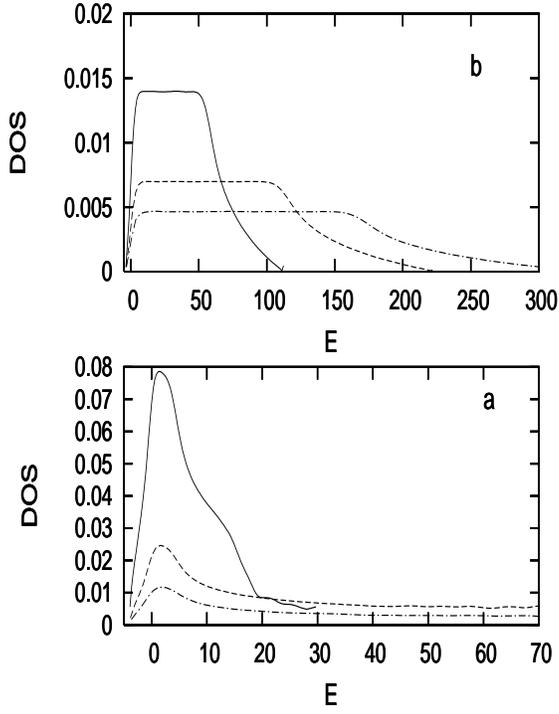}}}
\caption[]{
The DOS of a boson in a two-dimensional square lattice of size 150 $\times$ 150
in a quartic (a) and harmonic (b) potentials. 
In the bottom panel (a):
$q\,=\,0.25\times 10^{-5}$ (solid), $0.25\times 10^{-4}$ (dashes), and
$0.10\times 10^{-3}$ (dash-dot).
In top panel (b): $k\,=\,0.01$ (solid line), 0.02 (dashed line), 
and 0.03 (dash-dot). 
}
\end{figure}
\begin{figure}
\resizebox*{3.1in}{4.5in}{\rotatebox{270}{\includegraphics{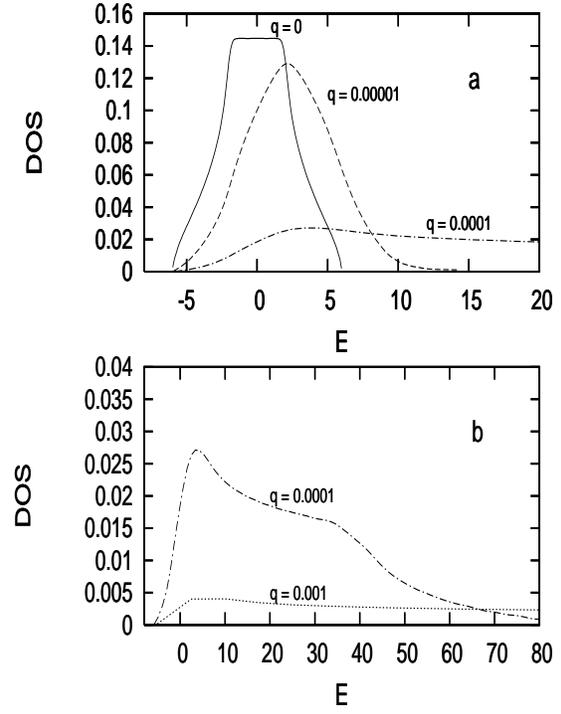}}}
\caption[]{
The effect of quartic potential strength on the DOS
of a boson in a three-dimensional cubic lattice of size 50 $\times$ 50
$\times$ 50.
}
\end{figure}
\begin{figure}
\resizebox*{3.1in}{2.5in}{\rotatebox{270}{\includegraphics{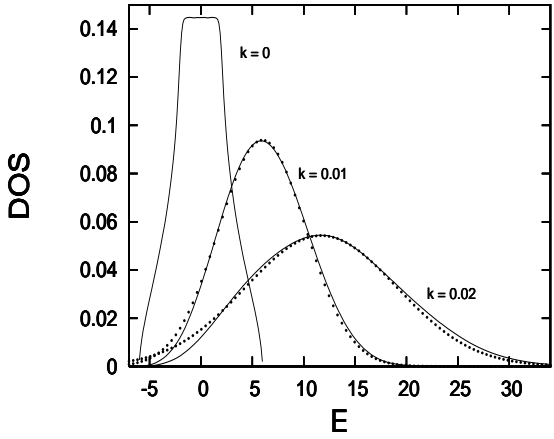}}}
\caption[]{
The effect of the strength of harmonic potential on DOS
of a boson in a three-dimensional cubic lattice of size 70 $\times$ 70
$\times$ 70 for the values of k shown on the curves. The dotted lines
are DOS's for an infinite dimensional lattice in the absence
of any confining potential, as discussed in the text.
}
\end{figure}
\begin{figure}
\resizebox*{3.1in}{2.5in}{\rotatebox{270}{\includegraphics{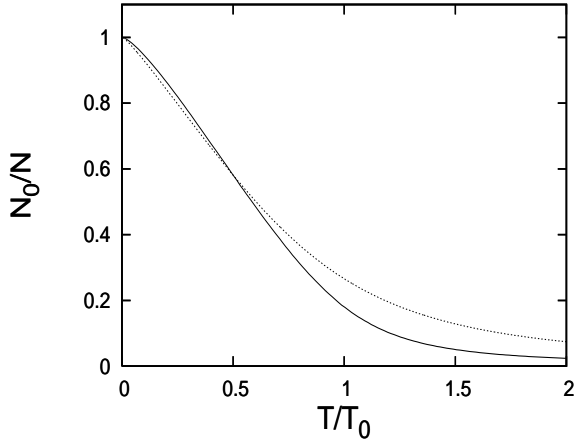}}}
\caption[]{
Temperature dependence of condensate fraction for bosons in 1d
optical lattices with confining potentials: lattice bosons
in a quartic trap of strength $q\, =\,0.25\times 10^{-6}$ (dots) and lattice
bosons in a harmonic trap of strength k = 0.001 (solid).
The lattice size used is 1000 and the number of bosons is 600.
The values of $k_BT_0$ for these cases are
3.23, and 6.773, respectively.
}
\end{figure}
\begin{figure}
\resizebox*{3.1in}{2.5in}{\rotatebox{270}{\includegraphics{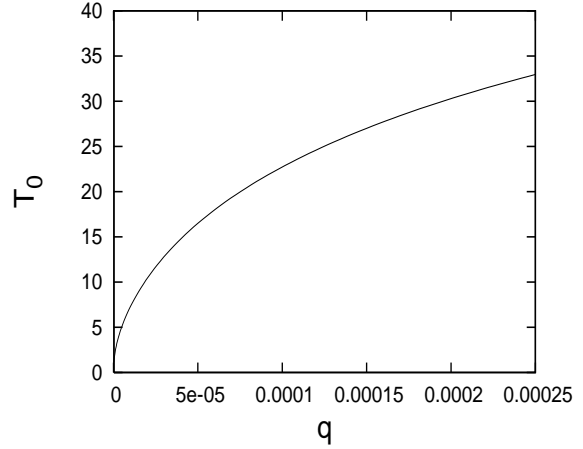}}}
\caption[]{
Quartic potential strength dependence of condensation temperature for
600 bosons in a 1d periodic lattice of size 1000. The lowest value
of q in this figure is $0.25\times 10^{-6}$.
}
\end{figure}
\begin{figure}
\resizebox*{3.1in}{2.5in}{\rotatebox{270}{\includegraphics{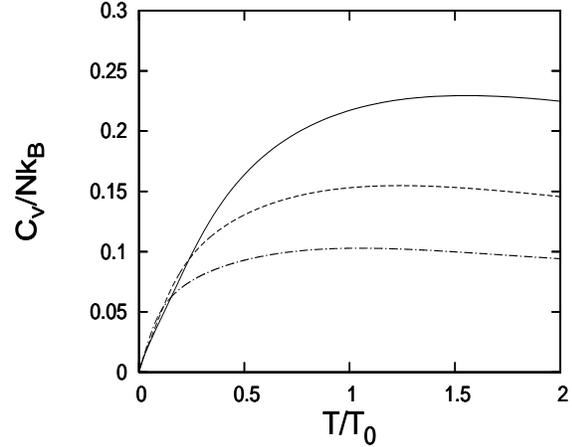}}}
\caption[]{
Temperature dependence of the specific heat of 600 bosons in
a 1d lattice of size 1000 in a quartic potential of stregnths
$q\,=\,0.25\times 10^{-6}$ (solid line), $0.25\times 10^{-5}$ (dots),
and $0.25\times 10^{-4}$ (dash-dot).
The values of $k_BT_0$ for these cases are
3.23, 6.91, and 14.98, respectively.
}
\end{figure}
\begin{figure}
\resizebox*{3.1in}{2.5in}{\rotatebox{270}{\includegraphics{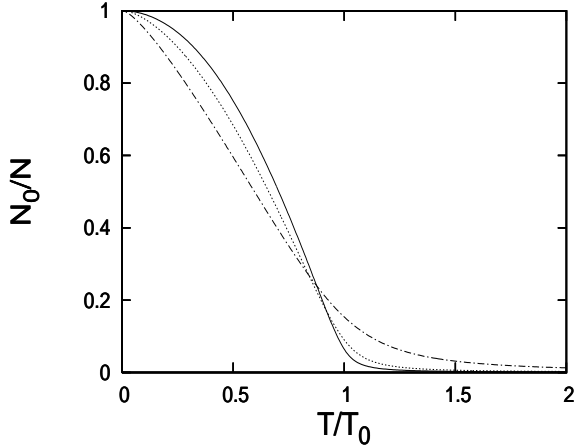}}}
\caption[]{
Comparison of the growth of condensate fraction of lattice
bosons in 2d: bosons in a 2d square lattice of size 100 $\times$ 100 
(dash-dot), lattice bosons in a quartic trap of 
strength $q\,=\,0.25\times 10^{-5}$ (dotted), and lattice
bosons in a harmonic trap of strength k = 0.01 (solid). The lattice
sizes used are 150 $\times$ 150 (harmonic trap case) and 100 $\times$ 100
(quartic trap case), and number of bosons is 600.
The values of $k_BT_0$ for these cases are
0.18, 1.187, and 3.19, respectively.
}
\end{figure}
\begin{figure}
\resizebox*{3.1in}{2.5in}{\rotatebox{270}{\includegraphics{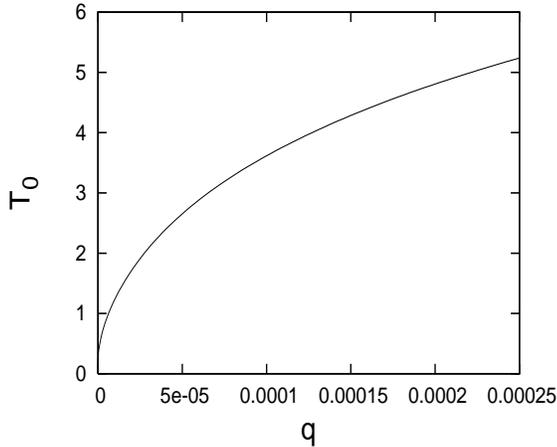}}}
\caption[]{
Quartic potential strength dependence of condensation temperature for
600 bosons in a 2d square lattice of size 100 $\times$ 100.
}
\end{figure}
\begin{figure}
\resizebox*{3.1in}{2.5in}{\rotatebox{270}{\includegraphics{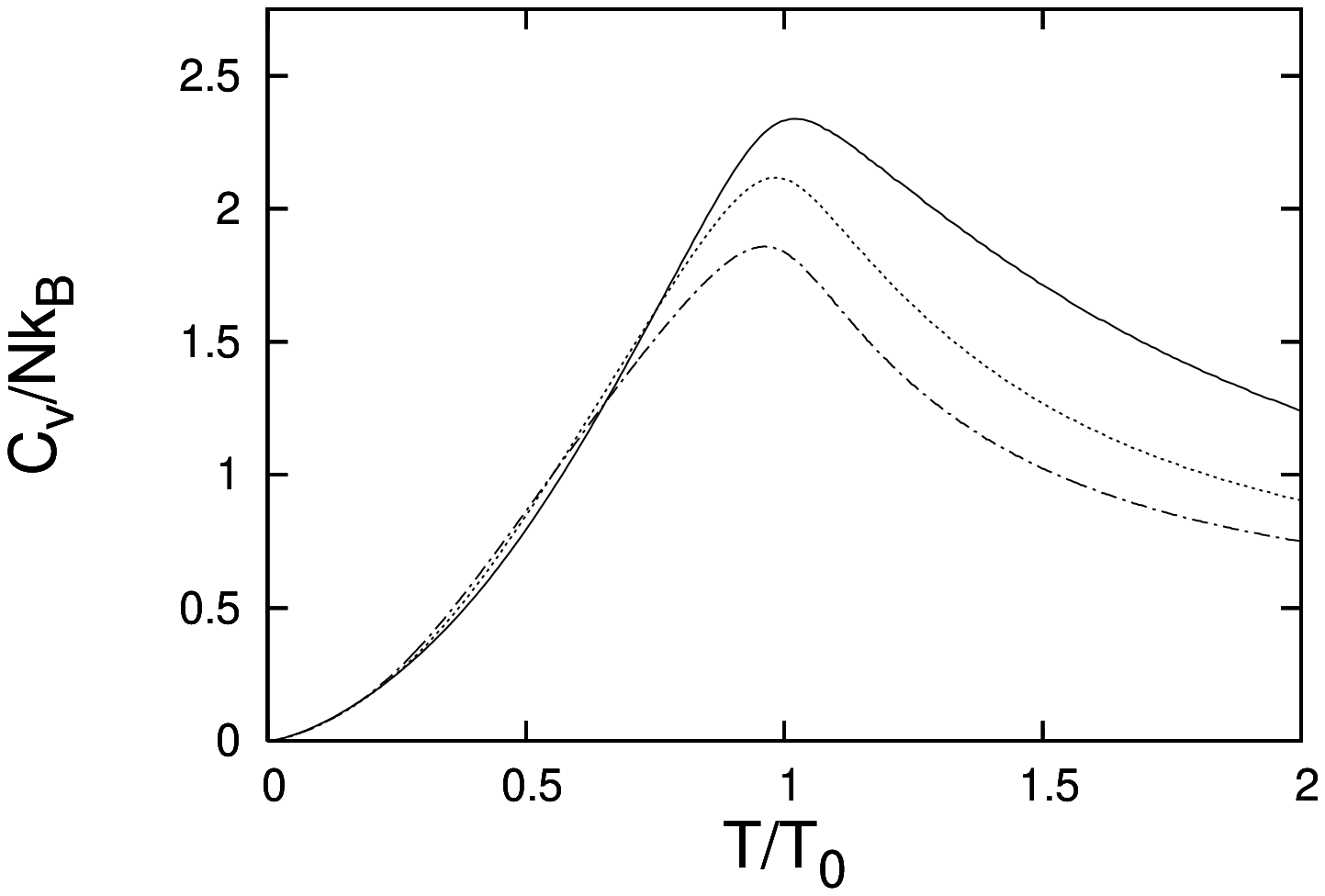}}}
\caption[]{
Temeperature dependence of specific heat of 600 bosons in
a 2d lattice of size 100 $\times$ 100 in a quartic potential of stregnths
$q\,=\,5\times 10^{-5}$ (solid line), $0.10\times 10^{-4}$ (dotted),
and $0.25\times 10^{-4}$ (dash-dot). The values of $k_BT_0$ for these cases are
1.187, 1.817, and 2.421, respectivley.
}
\end{figure}
\begin{figure}
\resizebox*{3.1in}{2.5in}{\rotatebox{270}{\includegraphics{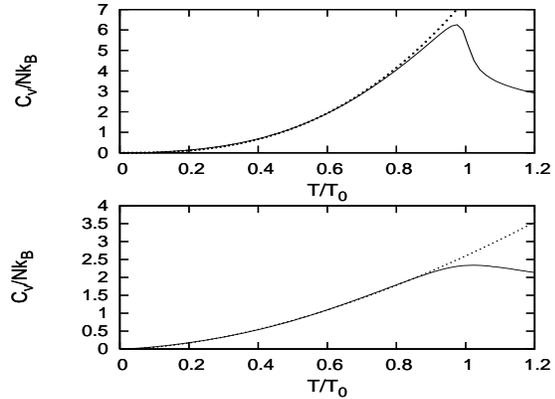}}}
\caption[]{
Fits to low temperature specific heats.
2d (bottom panel): the dotted line is a plot of $2.6\times (T/T_0)^{1.7}$.
3d (top panel): the dotted line is a plot of $7.5\times (T/T_0)^{2.65}$
}
\end{figure}
\begin{figure}
\resizebox*{3.1in}{2.5in}{\rotatebox{270}{\includegraphics{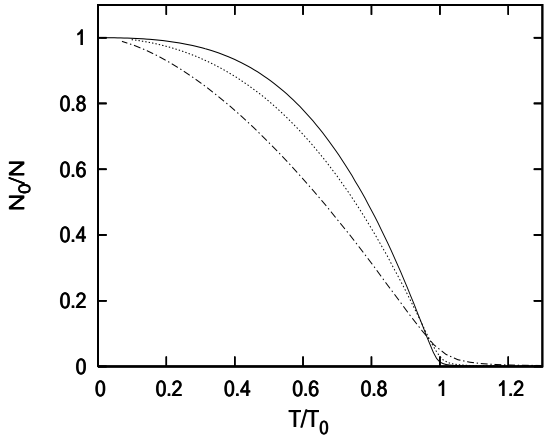}}}
\caption[]{
Comparison of the growth of condensate fraction of lattice
bosons in 3d: pure lattice bosons (dash-dot), lattice bosons in a
quartic trap of strength $q\,=\,10^{-5}$ (dotted), and
and lattice bosons in a harmonic trap of strength $k\,=\,0.01$ (solid). 
The lattice size used is 70 $\times$ 70 $\times$ 70 and number 
of bosons is 2500.
The values of $k_BT_0$ for these cases are
0.285, 1.23, and 2.08, respectively.

}
\end{figure}
\begin{figure}
\resizebox*{3.1in}{2.5in}{\rotatebox{270}{\includegraphics{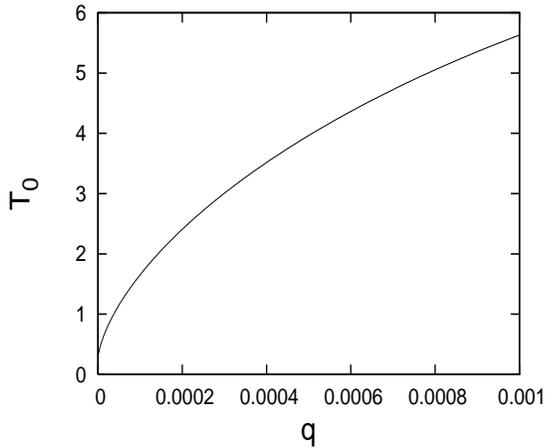}}}
\caption[]{
Quartic potential strength dependence of condensation temperature for
2500 bosons in a 3d square lattice of size 70 $\times$ 70 $\times$ 70.
}
\end{figure}
\begin{figure}
\resizebox*{3.1in}{2.5in}{\rotatebox{270}{\includegraphics{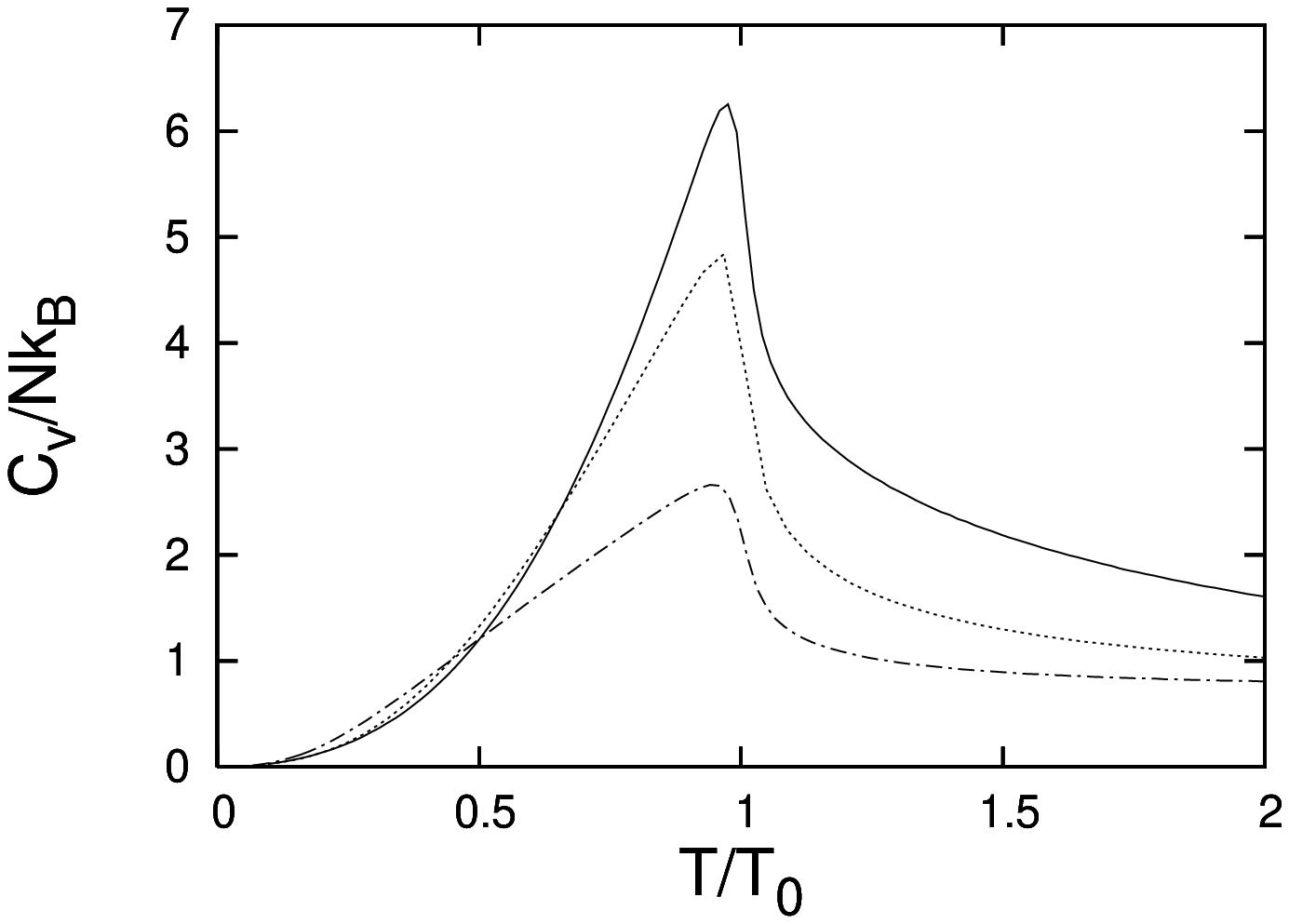}}}
\caption[]{
Temeperature dependence of specific heat of 2500 bosons in
a 3d lattice of size 70 $\times$ 70 $\times$ 70 in a quartic potential
of stregnths: $q\, =\, 10^{-5}$ (solid line), $10^{-4}$ (dots),
and $10^{-3}$ (dash-dot). The values of $k_BT_0$ for these cases are
1.23, 2.48, and 5.63, respectivley.
}
\end{figure}

\begin{thebibliography}{}
\bibitem{bloch}
I. Bloch, J. Dalibard, W. Zwerger, arXiv:0704.3011v1 [cond-mat.other].
\bibitem{greiner}
M. Greiner, O. Mandel, T. Esslinger, T. W. H\"{a}nsch, I. Bloch,
Nature {\bf 415}, 39 (2002).
\bibitem{stof}
T. St\"{o}ferle, H. Moritz, C. Schori, M. K\"{o}hl, T. Esslinger,
Phys. Rev. Lett. {\bf 92}, 130403 (2004).
\bibitem{schori}
C. Schori, T. St\"{o}ferle, M. Henning, M. K\"{o}hl, T. Esslinger,
Phys. Rev. Lett. {\bf 93}, 240402 (2004).
\bibitem{paredes}
B. Paredes, A. Widera, V. Murg, O. Mandel, S. F\"{o}lling, I. Cirac,
G. V. Shlyapnikov, T. W. H\"{a}nsch, I. Bloch, Nature {\bf 429},
277 (2004).
\bibitem{kohl}
M. K\"{o}hl, H. Moritz, T. St\"{o}ferle, C. Schori, T. Esslinger,
J. Low Temp. Phys. {\bf 138}, 635 (2005).
\bibitem{campbell}
G. K. Campbell, J. Mun, M. Boyd, P. Medley, A. E. Leanhardt, L. G. Maracassa,
D. E. Pritchard, W. Ketterle, Science {\bf 313}, 649 (2006).
\bibitem{folling}
S. F\"{o}lling, A. Widera, T. M\"{u}ller, F. Gerbier, I. Bloch,
Phys. Rev. Lett. {\bf 97}, 060403 (2006).
\bibitem{jaksch}
D. Jaksch, C. Bruder, J. I. Cirac, C. W. Gardiner, P. Zoller,
Phys. Rev. Lett. {\bf 81}, 3108 (1998).
\bibitem{zwer}
W. Zwerger, J. Opt. B: Quantum Semiclass. Opt. {\bf 5}, S9 (2003).
\bibitem{rey}
A. M. Rey, K. Burnett, R. Roth, M. Edwards, C. J. Williams, C. W. Clark,
J. Phys. B: At. Mol. Opt. Phys. {\bf 36}, 825 (2003).
\bibitem{pupilo1}
G. Pupillo, E. Tiesinga, C. J. Williams, Phys.Rev. A {\bf 68},
063604 (2003).
\bibitem{wessel}
S. Wessel, F. Alet, M. Troyer, G. G. Batrouni, Phys. Rev. A {\bf 70},
053615 (2004).
\bibitem{pollet}
L. Pollet, S. Rombouts, K. Heyde, J. Dukelsky, Phys. Rev. A {\bf 69},
043601 (2004).
\bibitem{giampaolo}
S. M. Giampaolo, F. Illuminati, G. Mazzarella, S. De Siena,
Phys. Rev. A {\bf 70}, 061601 (2004).
\bibitem{demarco}
B. DeMarco, C. Lannert, S. Vishveshwara, T.-C. Wei, Phys. Rev. A {\bf 71},
063601 (2005).
\bibitem{pupilo2}
G. Pupillo, A. M. Rey, G. G. Batrouni, Phys. Rev. A {\bf 74},
013601 (2006).
\bibitem{wild}
B. G. Wild, P. B. Blakie, D. A. W. Hutchinson, Phys. Rev. A {\bf 73},
023604 (2006).
\bibitem{murg}
V. Murg, F. Verstraete, J. I. Cirac, Phys. Rev. A {\bf 75}, 033605 (2007).
\bibitem{ramdassil}
R. Ramakumar, A. N. Das, S. Sil, Eur. Phys. J. D {\bf 42}, 309 (2007).
\bibitem{gygi}
O. Gygi, H. G. Katzgraber, M. Troyer, S. Wessel, G. G. Batrouni,
Phys. Rev. A {\bf 73}, 063606 (2006).
\bibitem{ramdas}
R. Ramakumar and A. N. Das, Phys. Rev. B {\bf 72}, 094301 (2005).
\bibitem{hooley}
C. Hooley and J. Quintanilla, Phys. Rev. Lett. {\bf 93}, 080404 (2004). 
\bibitem{metzner}
W. Metzner and D. Vollhardt, Phys. Rev. Lett. {\bf 62}, 324 (1989).
\bibitem{pethick}
C. J. Pethick and H. Smith, {\em Bose-Einstein
condensation in dilute gases} (Cambridge University Press, Cambridge, England, 2002).
\end{thebibliography}
\end{document}